\documentstyle[prl,aps]{revtex}
\def\build#1_#2^#3{\mathrel{
\mathop{\kern0pt#1}\limits_{#2}^{#3}}}

\def\fun#1#2{\lower3.6pt\vbox{\baselineskip0pt\lineskip.9pt
        \ialign{$\mathsurround=0pt#1\hfill##\hfil$\crcr#2\crcr\sim\crcr}}}

\def\bar{\overline}

\def\ben{\begin{equation}}
\def\be#1{\begin{equation}\label{eq:#1}}
\def\ee{\end{equation}}
\def\EC#1{(\ref{eq:#1})}

\begin{document}
  
\title{Self-Similar Approach to Violent Relaxation}
  
\author{Richard N. Henriksen and Lawrence M. Widrow}

\address{Department of Physics Queen's University, Kingston, Canada, K7L 3N6}

\maketitle

\begin{abstract} 

We consider the evolution of an initially cold, spherically symmetric
system of self-gravitating particles.  At early times the system
undergoes self-similar collapse of the type described by Fillmore \&
Goldreich and Bertschinger.  This stage of phase mixing soon gives way
to a period of violent relaxation driven by an instability in the
similarity solution.  The onset of violent relaxation is illustrated
by numerical simulations and supported by non-linear analysis of the
``scaled'' collisionless Boltzmann and Poisson equations.

\end{abstract}

\pacs{PACS numbers: }

The early evolution of an initially cold, self-gravitating system of
collisionless particles is governed by two dynamical processes: phase
mixing and violent relaxation [1].  With phase mixing, the initial,
single-velocity stream distribution function is wound into a tight
spiral pattern that can be described ultimately as a smooth (i.e.,
coarse-grained) function of the phase space variables ${\bf x}$ and
${\bf v}$.  With violent relaxation [2], the energies of individual
particles in the distribution function (DF) change as they move
through the time-dependent potential of the collapsing object.
Violent relaxation leads to a more rapid and complete mixing of the DF
and may erase all memory of initial conditions [2,3].

Clearly both phase mixing and violent relaxation operate during the
formation of gravitationally bound systems.  However the precise role
each plays in the relaxation process is largely unknown [4].  Moreover
it is not known whether realistic systems such as dark matter galactic
halos are truly mixed.  The question is important for dark matter
search experiments where the DF for the dark matter particles is
almost always treated as a smooth function of ${\bf x}$ and ${\bf v}$
(see however reference [5]).

Fillmore \& Goldreich [6] and Bertschinger [7] have found similarity
solutions which describe the collapse of initially cold, spherically
symmetric perturbations in an expanding universe.  While their
solutions exhibit some features of violent relaxation (particle
energies change as they move through a time-dependent potential) the
essential physics is that of phase-mixing.  In particular the
self-similar DFs found are characterized by a single thread in phase
space.

In this paper we demonstrate that the similarity solutions of [6,7]
are unstable and that the instability leads to true violent relaxation
of the system.  We begin by formulating the problem of self-similar
collapse in a novel way working directly with the DF rather than
particle orbits.  For simplicity we impose spherical symmetry and
treat only radial orbits.  The DF then depends on three variables;
radius, radial velocity, and time.  A scaling symmetry is imposed
which reduces, by one, the number of independent variables in the
collisionless Boltzmann equation (CBE).  The resultant equations are
characterized by a single dimensionless parameter which can be related
to the profile of the initial mass distribution.  Not surprisingly,
the characteristics of the CBE correspond to the orbit equations
derived in [6,7].  However the interpretation is different and leads
to a deeper understanding of the problem.  We next perform N-body
simulations of spherical radial collapse following the orbits of
particles (i.e., spherical shells) directly in phase space.  Early on
the DF exhibits characteristics of the self-similar solutions found in
[6,7].  However an instability soon develops leading to efficient
mixing of the DF.

The collisionless Boltzmann and Poisson equations for
a spherically symmetric system are

\be{sphericalv}
\partial_t f+v_r\partial_r f +\left({j^2\over r^3}-\partial_r\Phi\right)
\,(\partial_{v_r}f)=0~,
\ee

\be{sphericalp}
\partial_r(r^2\partial_r\Phi)~=~4\pi^2 \,\int\, dj^2
\int\,f(r,v_r,j^2)\,dv_r
\ee

\noindent where $f$ is the phase space mass density and $\Phi$ the is `mean field'
gravitational potential.  For purely radial orbits we introduce the
canonical distribution function $F(t,r,v_r)$ where $f\equiv \left
(4\pi r^2\right )^{-1}\delta(j^2)F$.  The equations then become:

\be{radialv}
\partial_t F+v_r\partial_r F-\partial_r\Phi~\partial_{v_r}F=0
\ee

\be{radialp}
\partial_r\left ( r^2\partial_r\Phi\right )~=~G\,\int\,F\, dv_r~.
\ee
 
We assume that a scaling symmetry exists along a direction $\bf k$
where [8,9]

\be{defk}
k^j\partial_j~=~ t\partial_t~+~\delta r \partial_r
~+~\nu v_r\partial_{v_r}
\ee

\noindent and $\delta$ and $\nu$ are real parameters.  It is convenient
to introduce an oblique time parameter $T(t)$ such that $k^j\partial_j
T=1$.  Self-similarity is imposed by requiring that all dimensional
quantities scale along ${\bf k}$ according to $A=A_o \exp{\beta T}$
where $\beta$ depends on $\delta$ and $\nu$ and on the dimensionality
of $A$.  Applying this procedure to the independent variables $r$ and
$v_r$ and to the dependent variables $F$ and $\Phi$ we find

\be{sscbe}
\nu\bar F+\left (Y-\delta X\right )
\frac{\partial \bar F}{\partial X}-
\left (\nu Y+\frac{d\bar\Phi}{dX}\right )
\frac{\partial \bar F}{\partial Y}=0
\ee

\be{sspoisson}
\frac{d}{dX}\left (X^2\frac{d\bar\Phi}{dX}\right )~=~\int \bar F dY
\ee

\noindent where $r = e^{\delta T}X$, $v_r = e^{\nu T}Y$, $F =
G^{-1} e^{\nu T}\bar F$, $\Phi = e^{2\nu T}\bar\Phi$, and $\nu =
\delta - 1$.  $\delta$ is set by initial conditions as illustrated in
the following simple example.  Consider a system which evolves from a
cloud of particles whose initial density distribution is $\rho =\lambda
r^{-\epsilon}$ where $\lambda$ and $\epsilon$ are constants.  We expect the
self-similar development to `remember' a kinematical constant
$G\lambda$ whose dimensions are given by $[G\lambda] = \left ({\it
length}\right )^\epsilon/\left ({\it time}\right )^2$.  Our general
solution `remembers' a kinematical constant $X\propto r/t^{\delta}$ as
this is invariant under the symmetry Eq.\,\EC{defk}.  This leads to the
correspondence $\epsilon = 2/\delta$.

Eqs.\,(6,7) are difficult to solve by direct methods
in part because of the complicated boundary conditions (e.g., the DF
should be continuous at $X\to 0$ as one passes from large negative
$Y$ (ingoing particles) to large positive $Y$ (outgoing particles)).
Instead we consider the characteristic curves of the
system:

\be{ode1}
\frac{dX}{ds}=Y- \delta X~;~~~
\frac{dY}{ds}=\left (1 - \delta\right )Y
-\frac{d\bar\Phi}{dX}~,
\ee

\be{ode3}
\frac{d\bar F}{ds}~=~\left (1 - \delta\right )\bar F
\ee

\noindent where $s$ is the path increment along a given curve.  
Eq.\,\EC{ode3} leads immediately to the result
$\bar F~=~e^{\left (1 - \delta\right )s}{\cal F}\left ( \zeta\right )$
where $\zeta\left (X,Y\right )$ is constant along a characteristic.
The differential equations for $X$ and $Y$ can be combined to obtain a
single second order differential equation

\be{odex}
\frac{d^2 X}{ds^2}~+~
\left (2\delta - 1\right )\frac{dX}{ds}~+~
\frac{d\bar\Phi_{\rm eff}}{X}~=~0
\ee

\noindent where $\bar\Phi_{\rm eff}\equiv\bar\Phi+\delta
\left (\delta - 1\right )X^2/2$.  While this equation is formally
equivalent to the orbit equations derived in [6,7] the interpretation
is different.  In particular the DF in [6,7] consist of particles that
lie along a single characteristic (i.e., $\delta-$\,function of
$\zeta$) as opposed to the smooth function $\bar F=\bar F(X,Y)$ given
by the solution to Eqs.\,(6,7).

The DFs found in [6,7] describe an eternal self-similar collapse
but say little about how the system enters or exits such a phase.
Moutarde et al.\,[10] have found features of the similarity solutions in
cosmological N-body simulations of initial overdensities concluding
that self-similar behavior of this type is ubiquitous in local
collapses.  Along different lines simple physical arguments [11]
as well as numerical simulations [12] can be
used to suggest analytic models for the final DF of violently-relaxed
galaxies.

Our numerical experiments set out to observe, directly in phase space,
the onset of self-similar behavior as well as the transition to a
final equilibrium configuration.  We model a spherically symmetric
distribution of particles traveling on purely radial orbits using a
shell code.  The initial density profile is characterized by a
power-law function of $r$; $\rho_i\propto r^{-\epsilon}$
($0<\epsilon<3$).  Potential difficulties at $r\to 0$ are treated by
using a softened force law $\propto M(r)/(r^2+r_0^2)$ where $M(r)$ is
the mass interior to $r$.  Alternatively we can place a reflecting
sphere at the origin or include a small amount of angular momentum.
Fig. 1 gives the phase space coordinates of all shells in the
simulation at six different times starting from an initial density
profile characterized by $\epsilon=3/2$.  For this experiment we set
$G=M(\infty)=1$ and use a softened force law with $r_0=0.01$.  Panels
b and c illustrate the essential behavior of the self-similar
solutions.  The distribution function in panel c is almost identical
to that of panel b provided we scale $r$ by a factor $s=1.8$ and $v_r$
by a factor $s^{\nu/\delta}=s^{1/4}$ where that latter scaling is set
by the relations $\delta=2/\epsilon$ and $\nu=\delta-1$.  Similar
results are found for an initial density profile $\rho\propto
r^{-5/2}$ where the velocity scale factor is $s^{-1/4}$ for a position
space scale factor $s$.  These results confirm the presence of
self-similarity as well as the connection established above between
the initial density profile and $\delta$.

Panels d and e illustrate new and unexpected behavior.  Evidently
there is an instability in the similarity solution.  Fluctuations
driven by this instability grow in time and quickly lead to mixing of
particles between neighboring streams erasing much of the fine-grained
spiral pattern.  Interestingly enough the pattern of instability is
replicated along the outer streams in the DF as one might expect given
the self-similar nature of the evolution.  We have checked that this
instability is real by varying the timestep by a factor of 100 and the
number of shells by a factor of 16 ($N=1000$ to $N=16,000$) and find
that the quantitative features of the instability remain the same.  In
addition very similar results arise in the $\rho_i\propto r^{-5/2}$
runs where a different shell spacing is used.  The onset of the
instability is sensitive to $r_0$ and can be delayed by making $r_0$
bigger.  Likewise adding a small amount of angular momentum will delay
growth of the fluctuations.  This is the expected result since violent
relaxation requires particles to move through a rapidly varying
potential which, for our example, is present at the origin.  Indeed the
equilibrium DFs constructed in [3,11] were based on this premise.

Our initial DF is finite in extent which leads to some interesting
effects.  In panel e the last particles have fallen through the center
and are now on their way toward apapsis.  However there are no more
particles falling in for the first time and so these last few
particles are less tightly bound than their earlier counterparts.  The
end of the distribution is therefore flung out to large radii like the
end of a whip.  Panel f shows the late time behavior of the system.
The phase space density at small $r$ is now quite smooth with no
evidence of the earlier fine-grained structure found in [6,7].  The
outer regions of the DF are characterized by a strikingly narrow
stream of particles.  These are the last few particles discussed
above.  Most are loosely bound and will eventually fall back into the
main distribution signaling a second round of secondary infall.

We find, in agreement with [10], that all cases where the initial
density law is shallower than $\epsilon=2$ have a density profile
during the self-similar phase $\propto r^{-2}$.  Alternatively, when
the initial density law is steeper than $\epsilon=2$, the self-similar
slope is equal to the initial slope.  The existence of two distinct
cases was noted originally in [6] and also in [10] and is discussed
below in the context of the non-linear dynamical analysis.

Eqs.\,\EC{ode1} yield the characteristics of a presumed smooth,
scaled, DF with $F$ along a characteristic given by the solution to
Eq.\,\EC{ode3}.  However the equivalent second order equation (14)
reveals that this system is not integrable unless the `friction' term
$(2\delta-1)dX/ds$ is either negligible or exactly zero.  The exactly
integrable case ($\delta =1/2$) corresponds to the condition that
the assumed self-similar symmetry (6) be an infinitesimal canonical
transformation generated by the radial action. Our dimensional
argument suggests that this case corresponds to the unphysical initial
density profile with $\epsilon =4$. In fact we show in a subsequent
paper that it is best associated with a distribution of particles
having constant angular momentum throughout.  For general $\delta$ the
friction term becomes negligible at small $X$.
 
When the non-integrability is manifest (the spiral `phase mixing'
orbits in phase space such as that shown in Figs. (1b,c)) the
functions $\zeta(X,Y)$ and $s_\zeta(X,Y)$ are not canonical
coordinates and can not be used to define a phase space volume element
everywhere. Consequently the DF is not defined over the whole phase
space. In the integrable limit ($X\to 0$) it becomes possible to
partition phase space with $\zeta$ (which becomes an `energy'
integral) and $s_\zeta$.  The evolution towards the existence of a
smooth DF parallels the numerical relaxation that we have observed in
Fig.\,1.
   
The other essential element for the existence of a smooth DF is coarse
graining. The fine-grained volume element must remain strictly zero in
the simulation, since the particles begin from rest. Nevertheless a
combination of round-off error and real instability leads to a
diffusion of this volume element throughout integrable phase space
(e.g. Fig.\,1f). This corresponds to the increase in entropy
associated with coarse grained DF's. The coarse graining formally
consists in taking the integrable characteristics to be dense in phase
space.
   
Fig.\,2 shows the topological structure of the characteristic curves
assuming the friction term is negligible.  For $\delta<1$ (the case
shown in the figure) there is a singular point at $Y=\delta X$ and
$d\overline\Phi_{eff}/dX=0$ i.e., simultaneously a turning point and
an extremum in the effective potential.  It is easy to show that this
is a saddle point.  At infinity there are two more singular points,
one stable node and one unstable node. These connect in a
straightforward way to the curves of Fig.\,2.  In this case
($\epsilon>2$) the numerical experiments indicate that the initial
density distribution is imported into the self-similar region.  When
$\delta>1$ the effective potential increases monotonically and there
are no singular points at finite X.  Moreover one finds that the
stable node at infinity is replaced by a saddle while the unstable
node remains unchanged. It is as though the region of
violent-relaxation extends to infinity.  The numerical experiments
show that in this case all memory of the initial density distribution
is lost in favor of a universal profile that is close to $r^{-2}$.
(Our analysis suggests that a universal profile $\rho(r)\propto
r^{-2}t^{-1}\left(\log{r/t^{1/2}}\right )^{-1/3}$ during the collapse
phase.  The logarithmic correction factor is reminiscent of what is
found in the equilibrium models [11].  At present our numerical
simulations do not have the dynamic range to verify this.)  Fillmore
\& Goldreich describe $\delta<1$ as the case where the inner halo is
dominated by particles with small apapsis and $\delta>1$ as the case
where particles are spread throughout the halo in agreement with the
discussion above.
   
The connection between the topology of the coarse-grained DF and the
ultimate density law in the self-similar region leads us to suggest
that {\it the appropriate pair of separatrices of the saddle point
form an effective boundary to the violently relaxed region of phase
space (and hence to the region of entropy increase)}. By contrast the
radial instability, which seems to be the principal mechanism for
relaxation in phase space, is present quite independently of the
topology of the DF characteristics.

Our results provide a simple yet dramatic example of the interplay
between phase mixing and violent relaxation.  The next step will be to
include angular momentum, deviations from spherical symmetry, and
small-scale clumpiness with the hope of better understanding the
formation and structure of realistic systems.

We are grateful to J. Dursi for help with the numerical simulations
and to M. Duncan, M. H. Lee, S. Tremaine, and G. Holder for valuable
conversations.  This work was supported in part by the Natural Science
and Engineering Research Council of Canada.

\bigskip

\bigskip

FIGURE 1: Evolution of a system of particles in phase space $(r,v_r)$
starting with zero velocity.  $G=M(\infty)=1$ which sets the units for
radius, time, and velocity.  The initial density profile is $\propto
r^{-3/2}$.  Panels a-f correspond to $t=0.0, 0.4, 0.6, 1.0, 1.2,
5.6\,$.

\bigskip

FIGURE 2: Phase diagram for the characteristic curves of the
self-similar CB and Poisson equations with $\delta = 0.8$ ($\epsilon =
5/2$).  The star marks the saddle point discussed in the text.  Inset:
Effective potential $\Phi_{\rm eff}$ as a function of $\log_{10}(X)$.
Solid and dotted curves are for $\epsilon=5/2$ and $3/2$ respectively.

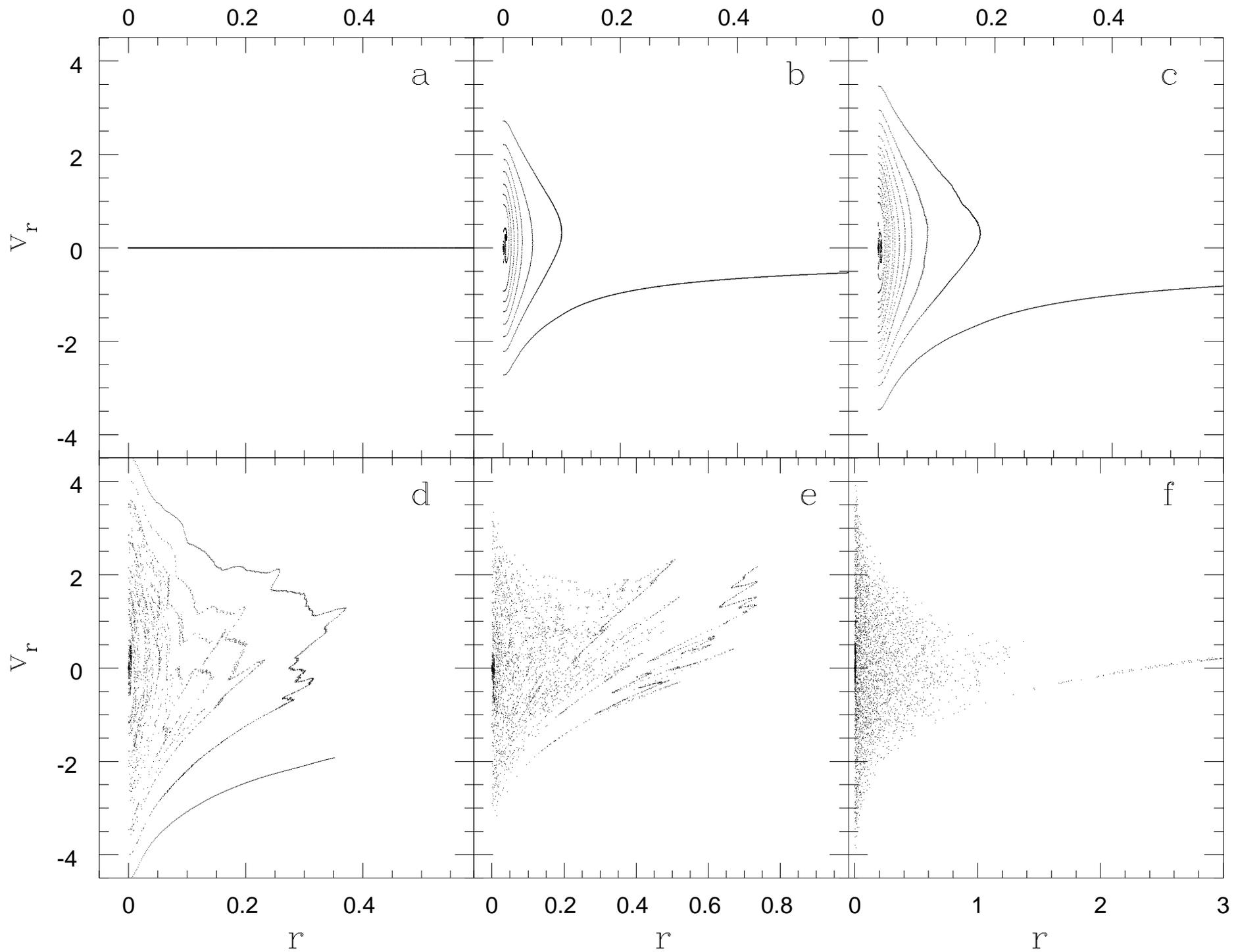
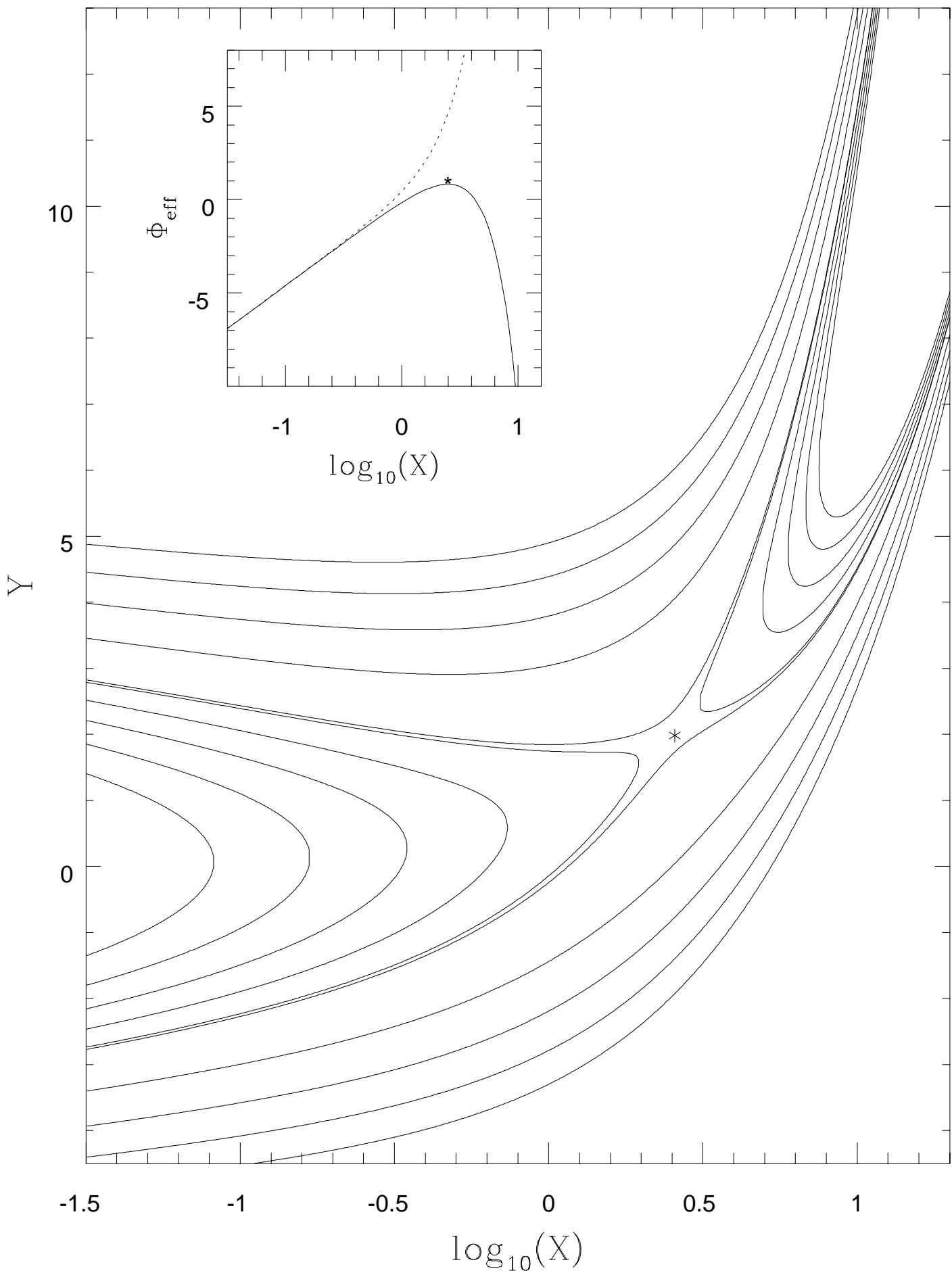
\end{document}